\documentclass{article}

\usepackage{algorithm}
\usepackage{algorithmic}
\usepackage{pifont}
\usepackage{amsmath}
\usepackage{color}
\usepackage{nomencl}
\usepackage{graphicx}
\usepackage{multirow}
\usepackage{booktabs}
\usepackage{bigstrut}
\usepackage{xcolor}
\usepackage{bbding}
\usepackage{graphicx}
\usepackage{CJK}
\usepackage{bm}
\usepackage{indentfirst}
\usepackage[colorlinks,linkcolor=black,anchorcolor=black,citecolor=black,urlcolor=black]{hyperref}
\usepackage{setspace}
\usepackage{makecell}
\usepackage{threeparttable}
\usepackage{arydshln}
\usepackage{multicol}
\usepackage{hhline}
\usepackage[misc]{ifsym}
\usepackage{balance}
\usepackage{url}

\usepackage{tabulary,graphicx,times,caption,fancyhdr,amsfonts,amssymb,amsbsy,latexsym,amsmath}
\usepackage[utf8]{inputenc}
\usepackage{url,multirow,morefloats,floatflt,cancel,tfrupee,textcomp,colortbl,xcolor,pifont}
\usepackage[nointegrals]{wasysym}
\usepackage{authblk}

\usepackage[paperheight=11in,paperwidth=8.3in,margin=2.5cm,headsep=.7cm,top=2.5cm]{geometry}

\usepackage[numbers,sort&compress]{natbib}

\usepackage{float,xcolor}

\title{NSSIA: A New Self-Sovereign Identity Scheme with Accountability}
\author[1]{Qiuyun Lyu}
\author[1]{Shaopeng Cheng}
\author[1]{Hao Li}
\author[2]{Junliang Liu}
\author[1]{Yanzhao Shen}
\author[1]{Zhen Wang}

\affil[1]{School of Cyberspace, Hangzhou Dianzi University, Hangzhou, Zhejiang 310018, China}
\affil[2]{Security department, Hangzhou Meichuang Technology Co., Ltd, Hangzhou, Zhejiang 310011, China}

\begin{document}
\maketitle

\begin{abstract}
		Self-Sovereign Identity (SSI) is a new distributed method for identity management, commonly used to address the problem that users are lack of control over their identities. 
		However, the excessive pursuit of self-sovereignty in the most existing SSI schemes hinders sanctions against attackers. 
		To deal with the malicious behavior, a few SSI schemes introduce accountability mechanisms, but they sacrifice users' privacy. 
		What's more, the digital identities (static strings or updatable chains) in the existing SSI schemes are as inputs to a third-party executable program (mobile app, smart contract, etc.) to achieve identity reading, storing and proving, users' self-sovereignty are weakened. 
		To solve the above problems, we present a new self-sovereign identity scheme to strike a balance between privacy and accountability and get rid of the dependence on the third-party program. 
		In our scheme, one and only individual-specific executable code is generated as a digital avatar-i for each human to interact with others in cyberspace without a third-party program, in which the embedding of biometrics enhances uniqueness and user control over their identity. 
		In addition, a joint accountability mechanism, which is based on the shamir (t, n) threshold algorithm and a consortium blockchain, is designed to restrict the power of each regulatory authority and protect users' privacy. 
		Finally, we analyze the security, SSI properties and conduct detailed experiments in term of the cost of computation, storage and blockchain gas. 
		The analysis results indicate that our scheme resists the known attacks and fulfills all the six SSI properties. 
		Compared with the state-of-the-art schemes, the extensive experiment results show that the cost is larger in server storage, blockchain storage and blockchain gas, but is still low enough for practical situations. 
\end{abstract}

\section{Introduction}\label{sec1}
Identity management (IdM) has experienced increased interest due to the ever growing demand for digital identities, as people become overly dependent on online services \cite{muhle2018survey}. 
However, each traditional IdM system usually adopts centralized authorization, authentication and maintains identity data independently \cite{liu2017identity}. 
As a result, enormous online IdM services force people to manage a large number of digital identities, which leads to the problem of identity fragmentation \cite{soltani2021survey} and is vulnerable to identity attacks, such as identity impersonation, privacy leakage, identity fraud, etc \cite{windley2021sovrin}. 
Even worse, users are lack of control and ownership over their digital identities in the traditional IdM \cite{dunphy2018first,zwitter2020digital}. 
Therefore, a distributed method for identity management called Self-Sovereign Identity (SSI) is proposed \cite{noauthor_path_nodate}, in which the users are central to the administration of identities. 
And, fortunately, the rise of distributed ledger technology (DLT), such as blockchain, has also made it possible to construct self-sovereign identities \cite{manski2020distributed,darnell20213,grech2021blockchain}. 
In comparison to the centralized management used by the traditional IdM, SSI schems shift decisions authority to users through secured DLT \cite{boysen2021decentralized} and allow them to possess full control of their identities and data \cite{toth2019self,kondova2020self,schardong2021self,wang2020self,freytsis2021development}. 

According to the goals to achieve, existing SSI schemes can be divided into the following three categories: \textit{junior SSI schemes} \cite{takemiya2018sora,zhou2019everssdi,niu2021self,westerkamp2019tawki,stokkink2018deployment}, \textit{SSI schemes with sybil-resistance} \cite{lee2019sims,zheng2019blockchain,hamer_private_nodate,othman2018horcrux,bandara2021blockchain}, and \textit{SSI schemes with accountability} \cite{stokkink2021truly,maram2021candid}. 
To give users' control over their identities and data, \textit{junior SSI schemes} adopt DID standard \cite{takemiya2018sora}, smart contracts \cite{zhou2019everssdi,niu2021self} or credential chain \cite{stokkink2018deployment}, etc. 
And static strings, such as DIDs, addresses of smart contracts, or updatable chains are employed to identify the users. 
However, the fact that users can hold as many identities as they want facilitates the implementation of sybil attacks. 
Therefore, many scholars introduced additional certificate authority \cite{lee2019sims} and biometrics \cite{othman2018horcrux,hamer_private_nodate,zheng2019blockchain,bandara2021blockchain} to ensure that each user has  one and only DID-based digital identity in their \textit{SSI schemes with sybil-resistance}. 
But unfortunately, the above schemes can not reveal the identities of malicious users. 
To deal with the problem, \textit{SSI schemes with accountability} \cite{stokkink2021truly,maram2021candid} are proposed. 
Since users are represented by credential chains in \cite{stokkink2021truly}, a regulatory authority checks the malicious credentials with the personal information in a central registry to identify the malicious users. 
But the audit of malicious users initiated by a single regulatory authority may lead to the serious problems of inadequate regulation or injustice. 
Different from the scheme \cite{stokkink2021truly}, the problems caused by a single regulation are overcome in the paper \cite{maram2021candid}. 
Specifically, the sanctions lists and a fuzzy matching method based on secure multi-party computation are applied to identify the credentials of suspicious users. 
However, both the central registry \cite{stokkink2021truly} and the sanctions lists \cite{maram2021candid} inevitably leak users privacy. 

In the other hand, metaverse, as the evolving paradigm of the next generation of the Internet \cite{wang2022survey}, will contain enormous amounts of applications and bring new challenges to the SSI. 
And, metaverse is considered as a massive virtual environment parallel to the physical world, in which users interact through digital avatars \cite{lee2021all}. 
That is, digital avatars are executable programs which own and control their identities for the user's physical self \cite{lee2021all,wang2022survey}. 
However, the digital identities (static strings or updatable chains) in the existing SSI schemes are all used as inputs to a third-party executable program (mobile app, smart contrat, etc.) to achieve identity reading, storing and proving. 
Thus, the existing SSI schemes cannot play well in metaverse and also weaken users' self-sovereignty. 
In detail, the dependence on a third-party executable program during the usage of SSI inevitably leads to the problems of single point of failure and privacy leakage. 

Inspired by the digital avatars in metaverse, and taking the above problems in the existing SSI schemes into account, we propose a new self-sovereign identity scheme with accountability. 
And the contributions of the proposed scheme are summarized as follows:  
\begin{itemize}
  \item[(i)] We propose a new self-sovereign identity scheme with accountability (NSSIA), in which executable code is introduced to allow users to control their identities completely and the balance between privacy and accountability is achieved. 
  \item[(ii)] To get rid of the dependence on the third-party programs, one and only individual-specific executable code is distributed to each user, where the user's biometrics are embedded to enhance uniqueness and user control. The hash of the executable code is used as an identifier and each user can use his/her own local executable code to store, read, and prove identities with network servers. 
  For simplicity, the term "digital avatar" in metaverse is borrowed and reformed to "digital avatar-i" to denote the executable code focusing on digital identity. 
  \item[(iii)] In order to regulate malicious users fairly without violating privacy, a joint accountability mechanism is introduced to decentralize the power of regulatory authorities and hide users' information in reality through shamir(t, n) threshold signature algorithm, while the impartial audit is further guaranteed by a consortium blockchain. 
  \item[(iv)] We analyze the proposed scheme in detail in terms of security, SSI properties in generation phase and conduct extensive experiments in the cost of computation, storage and blockchain gas. 
\end{itemize}

The rest of this article is organized as follows. 
Section \ref{sec2} introduces the related work of SSI schemes. 
In Section \ref{sec3}, the system model, security model and design goals are introduced and Section \ref{sec4} describes our scheme in detail. 
We analyze our proposed scheme in terms of security and performance in Section \ref{sec5} and Section \ref{sec6} respectively. 
Finally, the conclusion and future work are given in Section \ref{sec7}. 

\section{Related Work}\label{sec2}
\subsection{Junior SSI schemes} \label{subsec2.1}
Junior SSI schemes \cite{takemiya2018sora,zhou2019everssdi,niu2021self,westerkamp2019tawki,stokkink2018deployment}, are first proposed to allow users to control their own identities. 
To enable users to have full control over their identities, Takemiya et al. \cite{takemiya2018sora} designed a security protocol for storing encrypted personal information based on Hyperledger Iroha. 
The decentralized identifier (DID) \cite{noauthor_decentralized_nodate} was used as the unique identifier of each user, while entries that characterized a user's identity were represented in the form of verifiable claims \cite{VC}. 
For self-sovereignty, all the claims were stored locally on user's phone in encrypted form. 
Different from Takemiya et al. \cite{takemiya2018sora}, smart contracts were used to represent the user's identity in the paper of \cite{zhou2019everssdi} and \cite{niu2021self}. 
Concretely, they both designed a kind of smart contracts with addresses as identifiers, specifically for managing identities. 
Once published, these contracts were owned by the corresponding users. 
And, the user's identity information was stored in IPFS \cite{zhou2019everssdi} and stored in the user's device in the form of a Merkle tree \cite{niu2021self} for self-sovereignty. 
However, both DIDs and smart contract addresses are machine-readable static strings, which are difficult for users to understand, leading to the dilemma of managing digital identities. 

Then, a decentralized service architecture for self-sovereign social communication, proposed by Westerkamp et al. \cite{westerkamp2019tawki}, solves the above problem. 
In this scheme, the user's identifier was represented as a human-readable name which was generated by the smart-contract based Ethereum Name Service (ENS). 
Besides, the user's data was stored in his/her own API server, and the Uniform Resource Identifier (URI) of the server was stored and linked to the human-readable name on the blockchain. 
However, such human-readable identifier is still inherently static, which are easily impersonated by malicious users during use. 

Fortunately, this problem can be alleviated by a general provable claim model proposed in the paper of \cite{stokkink2018deployment}. 
With reference to the structure of the blockchain, the self-sovereign identity was designed as a growing chain of user's claims. 
And, the user's identity could be used only after the authentication of the verifier on the existing claims, thus alleviating the risk of identity being impersonated. 
But, due to the lack of necessary authentication before identity registration, users can create as many identities as they want, which facilitates the implementation of sybil attacks. 
\subsection{SSI schemes with Sybil-resistance}\label{subsec2.2}
In order to let each user have one and only digital identity, SSI schemes with sybil-resistance \cite{lee2019sims,zheng2019blockchain,hamer_private_nodate,othman2018horcrux,bandara2021blockchain} are proposed. 
A commitment scheme combined with zk-SNARK were introduced in \cite{lee2019sims} to provide integrity and privacy of user information simultaneously. 
In this scheme, to ensure integrity and avoid reuse, only after the user's information and the corresponding commitment were confirmed by the CA, a certificate would be issued to the user. 
And during usage, user's data was encrypted by zk-SNARK to prevent privacy leakage. 
However, the verification of the commitment by the CA can only guarantee the integrity of the information from the user, not the authenticity, which means the CA can be deceived by false information. 

For the authenticity and reliability of user identity, biometric identification is introduced into SSI schemes \cite{zheng2019blockchain,hamer_private_nodate,othman2018horcrux,bandara2021blockchain}. 
In 2018, Othman et al. \cite{othman2018horcrux} designed a novel method for decentralized biometric-based self-sovereign identity. 
In this scheme, the user's identity was created based on the DID specification. 
Also, in order to associate each user with their own identity, biometrics (fingerprint, face, voice, etc.) were encrypted and stored in the corresponding DID document. 
But unfortunately, biometric information is only collected through the user's mobile app, which presents an opportunity for adversaries to commit identity fraud. 

Then, in 2019, Hamer et al. \cite{hamer_private_nodate} proposed a unique self-sovereign identity management scheme to deal with the above problem. 
A user's biometrics were authenticated by the trusted organization to make sure that the user did have these biometrics. 
Besides, the collected biometrics were encrypted with a homomorphic signature algorithm to ensure that a user could't enroll twice in the system. 
But all the user's behaviors can be linked to the same digital identity, which leaks the user's privacy. 

A blockchain-based privacy protection unified identity authentication scheme is proposed in \cite{zheng2019blockchain}. 
In this scheme, the server would authenticate user information by online face verification using photos from a central database. 
In addition, a set of key derivation algorithms were designed to ensure the unlinkability between identity attribute information. 
However, the way a central database stores user information weakens users' control over their identities. 

In 2021, Bandara et al. \cite{bandara2021blockchain} proposed a blockchain and self-sovereign identity empowered digital identity platform. 
For full control over the data, the information required for registration (name, address, photo, etc.) submitted by the user was stored locally on the mobile device. 
And only with the user's consent, these information would be sent to the service provider (SP). 
Additionally, verification of information is achieved by comparison with physical documents, eliminating the need for SPs to store information. 

In a word, strict identity authentication, especially the introduction of biometrics, ensures that each user has one and only digital identity, effectively resisting sybil attacks. 
However, none of these schemes design accountability mechanisms to regulate malicious users who disrupt the order of the network. 
\subsection{SSI schemes with Accountability}\label{subsec2.3}
For the purpose of maintaining order in cyberspace, SSI schemes with accountability \cite{stokkink2021truly,maram2021candid} are proposed. 
In 2021, Stokkink et al. \cite{stokkink2021truly} designed a truly self-sovereign identity system based on Pedersen commitments, where the digital identities were implemented as data structures that held a list of credentials. 
And, for self-sovereignty, these data structures were stored on the users' devices. 
In terms of accountability, credential verifiers were required to keep audit logs, which were actually composed of credentials presented by users. 
Then, malicious users could be identified by a single regulatory authority through analyzing audit logs and comparing them with the personal information in a central registry. 
But, several problems such as inadequate regulation and injustice may arise due to the reliance on a single regulatory authority. 

Also in 2021, Maram et al. \cite{maram2021candid} presented a decentralized identity management with legacy compatibility, sybil-resistance and accountability. 
Instead of an additional credential issuer, all credentials characterizing the user's identity in this scheme were imported from existing web service providers. 
And a deduplication protocol based on secure multi-party computation (MPC) was designed to prevent the reuse of these credentials. 
Besides, in order to address the drawbacks of the single regulatory authority, an MPC-based fuzzy matching method was proposed, which can find the digital identities of the corresponding malicious users according to the sanctions lists. 
However, legacy compatibility does not change the status quo of data stored by existing web service providers, which remains out of the user's control. 
In addition, both the central registry and the sanctions lists introduced in the above schemes to regulate malicious behavior inevitably sacrifice users privacy. 
\section{Models and Design Goals} \label{sec3}
\subsection{System model}\label{subsec3.1}
\begin{figure*}[!htbp]
    \begin{center}
        \includegraphics[width=\linewidth]{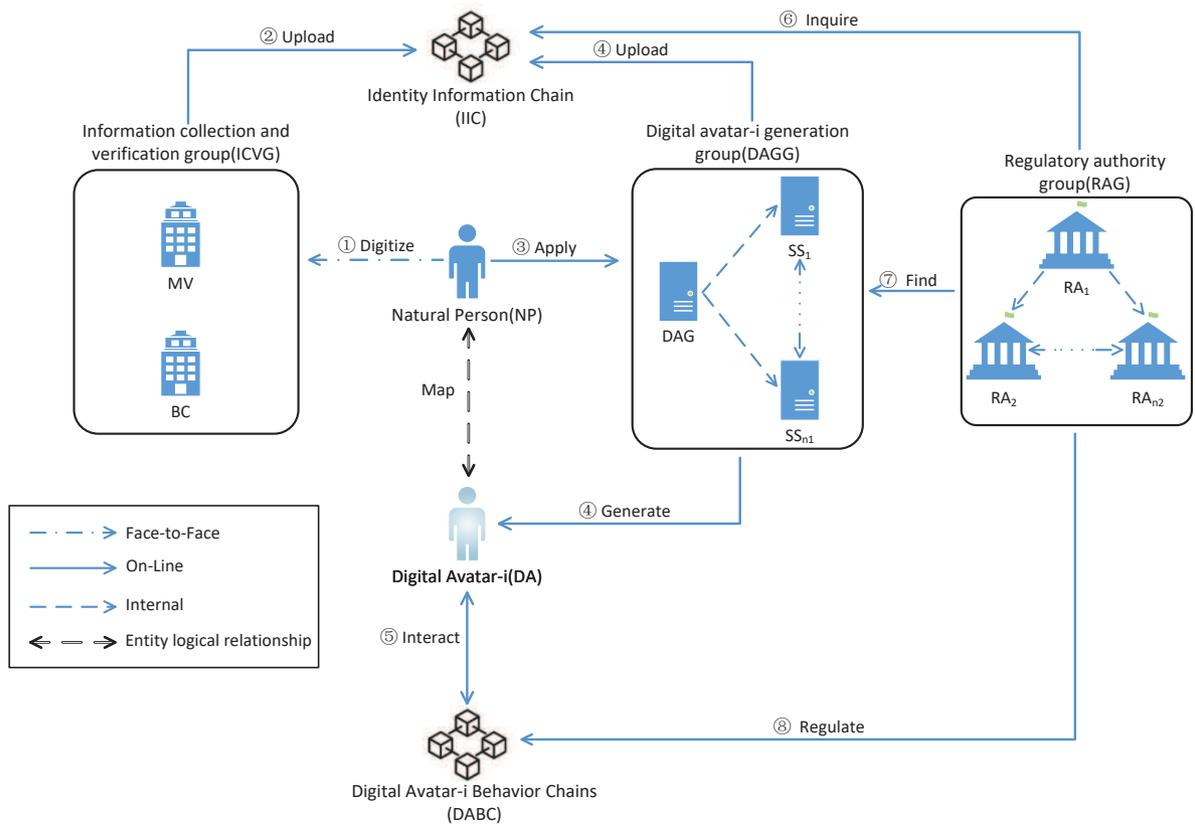}
    \end{center}
    \caption{System model\label{fig1}}
\end{figure*}
Our system model consists of seven entities, as Figure~\ref{fig1} shows, a natural person (NP), a digital avatar-i (DA), two blockchains: an identity information chain (IIC), a digital avatar-i behavior chain (DABC) and three groups: an information collection and verification group (ICVG), a digital avatar-i generation group (DAGG), and a regulatory authority group (RAG). 

\begin{enumerate}
    \item[\qquad $\bullet$] \textbf{NP}, Natural Person, refers to a person living in the physical world. 
    He/She can digitize himself/herself through the ICVG and apply to the DAGG for a DA. 
    \item[\qquad $\bullet$] \textbf{DA}, Digital Avatar-i, is an individual-specific executable program focusing on the identity dimension of the digital avatar, which stands for a living person to interact with others in cyberspce, and has one-to-one relationship with NP. 
    \item[\qquad $\bullet$] \textbf{ICVG}, Information Collection and Verification Group, validates that the requestor is one and only breathing person in the physical world and provides digitalizing service for him/her. It contains two types of entities, namely metadata verifier (MV) and biometric collector (BC). MV proves the requestor's existence in physical space through metadata, such as name, identity number and address, etc. BC collects two types of distinct biometric data, where one is as a permanent proof and the other is for activating the DA. 
    \item[\qquad $\bullet$] \textbf{IIC}, Identity Information Chain, is a consortium blockchain. It is mainly responsible for recording the proof of physical identity information (metadata and biometric data), the hash of DA, and making sure each NP has only one proof. 
    \item[\qquad $\bullet$] \textbf{DAGG}, Digital Avatar-i Generation Group, generates a unique DA for each NP. It contains two types of entities, namely digital avatar-i generator (DAG) and secure storages (SSs). At first, DAG verifies the identity of the applicant with the data in the IIC, and then generates the sole DA for him/her. SSs, which contain SS$_1 \cdots $ SS$_n$, use shamir(t, n) threshold algorithm to safely store the metadata of NP and the hash of DA.
    \item[\qquad $\bullet$] \textbf{DABC}, Digital Avatar-i Behavior Chain, is an infrastructure which is composed of multiple blockchains, supporting all kinds of decentralized applications (Dapps). These Dapps provide services for DAs in cyberspace and the DABC keeps their historical records for accountability. 
    \item[\qquad $\bullet$] \textbf{RAG}, Regulatory Authority Group, is responsible for regulating NP by monitoring the DA's activities in the DABC. 
    And it is composed of n regulatory authorities (RA$_1$ $\cdots$ RA$_n$), where at least three of them can hold suspicious users accountable. 
\end{enumerate}

In order to securely and privately take part in various activities such as work, study and entertainment in cyberspace, especially the metaverse, a NP needs to map himself/herself to one DA. 
In detail, there are four steps to achieve the mapping. 
Firstly, the NP needs to digitize himself/herself through the ICVG, where MV verifies the descriptive metadata of the NP and BC collects the NP's biometric data. 
Secondly, the ICVG uploads the proof information to the IIC and replies the NP with a certificate. 
Thirdly, the NP applies to the DAGG for a DA with the certificate. 
At last, the DAGG verifies the authenticity of the NP by checking whether the live biometric data matches the proof in the IIC. 
If the verification is passed, the DAGG generates the DA and uploads the hash of the DA to the IIC. 
Afterwards, the NP uses the corresponding DA to live in the cyberspace without a third-party program. 

It is worth noting that, once there is a malicious DA, the RAG can map him/her to the corresponding NP throuth inquiring the IIC and finding the metadata in DAGG. 
For constructing a safe and orderly cyberspace, a DA is supposed to interact with the Dapps based on DABC. 
In this way, the RAG can regulate a malicious NP through monitoring the DA's historical and future behaviors. 
\subsection{Security model}\label{subsec3.2}
In NSSIA, we have the following security assumptions.
	\begin{itemize}
		\renewcommand{\labelenumii}{(\arabic{enumii})}
		\item[$\bullet$] An adversary can monitor, intercept, modify and insert the messages into the public channel \citep{dolev1983security}. 
		And he/she can breach no more than half of the entities in each group of ICVG, DAGG and RAG within a certain period of time. 
		\item[$\bullet$] A NP and a DA are considered as malicious entities. 
		A NP would submit false information or fraudulently use anyone's information, and a DA can be modified or illegally used by an adversary in the cyberspace. 
		\item[$\bullet$] Entities in the ICVG, DAGG and RAG are regarded as semi-honest. 
		They will perform the protocol strictly but are curious about the information. 		
		\item[$\bullet$] The DABC is a semi-honest entity. It is a public chain which is secured by consensus algorithms, and the miners perform the protocol strictly but are curious about the information. 
		The IIC is a trusted consortium chain which is jointly managed by the ICVG, DAGG and RAG. 
		\item[$\bullet$] We assume that the standard cryptographic algorithm used in our scheme is secure and unbreakable.
	\end{itemize}
    \subsection{Design Goals}\label{subsec3.3}
    According to the aforementioned system model and security model, the design goals of our scheme are as follows.
	\begin{enumerate}
	 	\item[(i)] User friendly: A user (NP) accesses a service in cyberspace with a digital avatar-i (DA) in a convenient way and the user owns and controls it. 
		\item[(ii)] One-to-one: In order to build an orderly cyberspace, a user (NP) has one and only digital avatar-i (DA). 
		And all the behaviors of the DA belong to the only one NP. 
		\item[(iii)] Linkability with condition: For the security of cyberspace and the privacy of a user (NP), the identity mapping (the NP and the DA) is encrypted and stored in a distributed way (different piece of it in each SS$_i$). 
		Only three or more of the RAs can jointly decrypt it and recover the detail of identity. 
	\end{enumerate}
\section{Proposed NSSIA}\label{sec4}
The NSSIA generates a unique digital avatar-i for a user to ensure his/her conditional identity privacy when interacting with each others in cyberspace, especially the metaverse. 
Concretely, the NSSIA is mainly divided into five phases. 
The first phase initializes the entities in the IIC, ICVG, DAGG and RAG to generate their keys. 
The second phase lets a NP digitize himself/herself through the ICVG, where the ICVG verifies the authenticity of NP's metadata, collects NP's biometric data and writes the proof information to the IIC, as shown in steps \ding{192}-\ding{193} of the Figure~\ref{fig1}. 
And in the third phase, the NP applies to the DAGG for a DA in which the DAGG checks the metadata of the NP, generates a DA and records the DA generation transaction to the IIC, as shown in steps \ding{194}-\ding{195} of the Figure~\ref{fig1}. 
The NP can use his/her own DA to interact with the Dapps built on DABC in the fourth phase, as shown in steps \ding{196} of the Figure~\ref{fig1}. 
Lastly, the RAG regulates a malicious NP with the mapping DA's behaviors in the DABC and the data in the IIC and the DAGG, as shown in steps \ding{197}-\ding{199} of the Figure~\ref{fig1}. 
To elaborate the NSSIA clearly, we give the notations used in our scheme in Table~\ref{tab1}.
\begin{table}[!htbp]
	\centering
	\caption{Notations used in our scheme\label{tab1}}
	\begin{threeparttable}
		\def\arraystretch{1.15}
		\begin{tabular}{ll}
			\toprule[1.5pt]
			Notations &Description\\
			\midrule[0.8pt]
			$MK$ & Master Key\\
			$PK_e$ & Public key of entity $e$\\
			$SK_e$ & Secret key of entity $e$ \\
			$SubK_e$ & Subkey of entity $e$\\
			$ESubK_e$ & Encrypted subkey of entity $e$ \\  
			$n_1,\ t_1$ & \makecell[l]{The number of secure storages is $n_1$ and \\the corresponding threshold is $t_1$}   \\
			$n_2,\ t_2$ & \makecell[l]{The number of regulatory authorities is\\ $n_2$ and the corresponding threshold is $t_2$}  \\         
			$a\oplus b$ &XOR operation of $a$ and $b$\\
			$H(\bullet)$ & Hash operation on $\bullet$\\
			$En(a, b)$ & Use $b$ to encrypt $a$\\
			$De(a,b)$ & Use $b$ to decrypt $a$ \\
			$Sig(a,b)$ & Use $b$ to sign $a$\\
			$Ver(a,b)$ & Use $b$ to verify $a$ \\
			$SecInfo$ & Encrypted identity information\\
			\bottomrule[1.5pt]
		\end{tabular}            
	\end{threeparttable}
\end{table}
\subsection{Initialization}\label{subsec4.1}
The IIC performs initialization to generate the public parameters, the master key (MK) and the corresponding subkeys (SubKs). 
In addtion, the entities in RAG, DAGG and ICVG generate their public and private keys. 
\subsubsection{IIC Initialization}\label{subsubsec4.1.1}
The IIC performs initialization to generate the public parameters, the $MK$ and the $SubK_e$, where the $SubK_e$ is the subkey of the entity $e$. 
In detail, it firstly selects a large prime $p$, an elliptic curve $E_p(a,b)$ and a base point $G$ with order $n$ under the finite field $F_p$. 
Then, it publishes the public parameters $P = \{p,E_p(a,b),G,n\}$ to the genesis block. 
Afterwards, it randomly selects a 128-bit AES key as the $MK$. 
Lastly, the IIC uses the shamir (t,n) threshold secret sharing algorithm \cite{shamir1979share} to generate the $SubK_e$ for each entities of DAGG and RAG. And we assume that the number of SSs and RAs is $n_1$ and $n_2$, and the corresponding thresholds are $t_1$ and $t_2$. 
Here are the details below. 

The IIC firstly chooses two polynomials of degree $t_{1}-1$ and $t_{2}-1$ shown as Equations (\ref{eq1}) and (\ref{eq2}), where $a_1, ..., a_{t_{1}-1} $, $b_1, ..., b_{t_{2}-1}$ are random numbers, and $N1$, $N2$ are bigger than each coefficient. 
\begin{equation}
    F_{SS}(x)=MK+a_1x+\cdots+a_{t_{1}-1}x^{t_{1}-1}mod(N1)\label{eq1}
\end{equation}
\begin{equation}
    F_{RA}(x)=MK+b_1x+\cdots+b_{t_{2}-1}x^{t_{2}-1}mod(N2)\label{eq2}
\end{equation}	

Next, the IIC chooses random numbers $x_i, i=1,2,..., n_1+n_2$, and substitutes the $x_i$ into Equations (\ref{eq1}) and (\ref{eq2}) to calculate the $n_{1}$ $SubK_{SS}s$ and $n_{2}$ $SubK_{RA}s$ ($n_{k}=2\times t_{k}-1,\ k=1,2$). 
\subsubsection{SS and RA Initialization}\label{subsubsec4.1.2}
The SS and RA use the $P$ published by IIC to generate their own public and private keys, referred to as $PK_{SS}/SK_{SS}$ and $PK_{RA}/SK_{RA}$, and publish the public keys. 
Then, the encrypted subkeys with $PK_{SS}$ and $PK_{RA}$ are obtained from the IIC by the SS and RA. Next, they decrypt the encrypted subkeys to recover the $SubK_{SS}$ and $SubK_{RA}$. 
\subsubsection{MV, BC and DAG Initialization}\label{subsubsec4.1.3}
MV, BC and DAG use the $P$ published by IIC to generate their own public and private keys, referred to as $PK_{MV}/SK_{MV}$, $PK_{BC}/SK_{BC}$ and $PK_{DAG}/SK_{DAG}$. 
\subsection{Digitization}\label{subsec4.2}
\begin{figure}[!htbp]
    \centering
    \begin{center}
        \includegraphics[width=1.0\columnwidth]{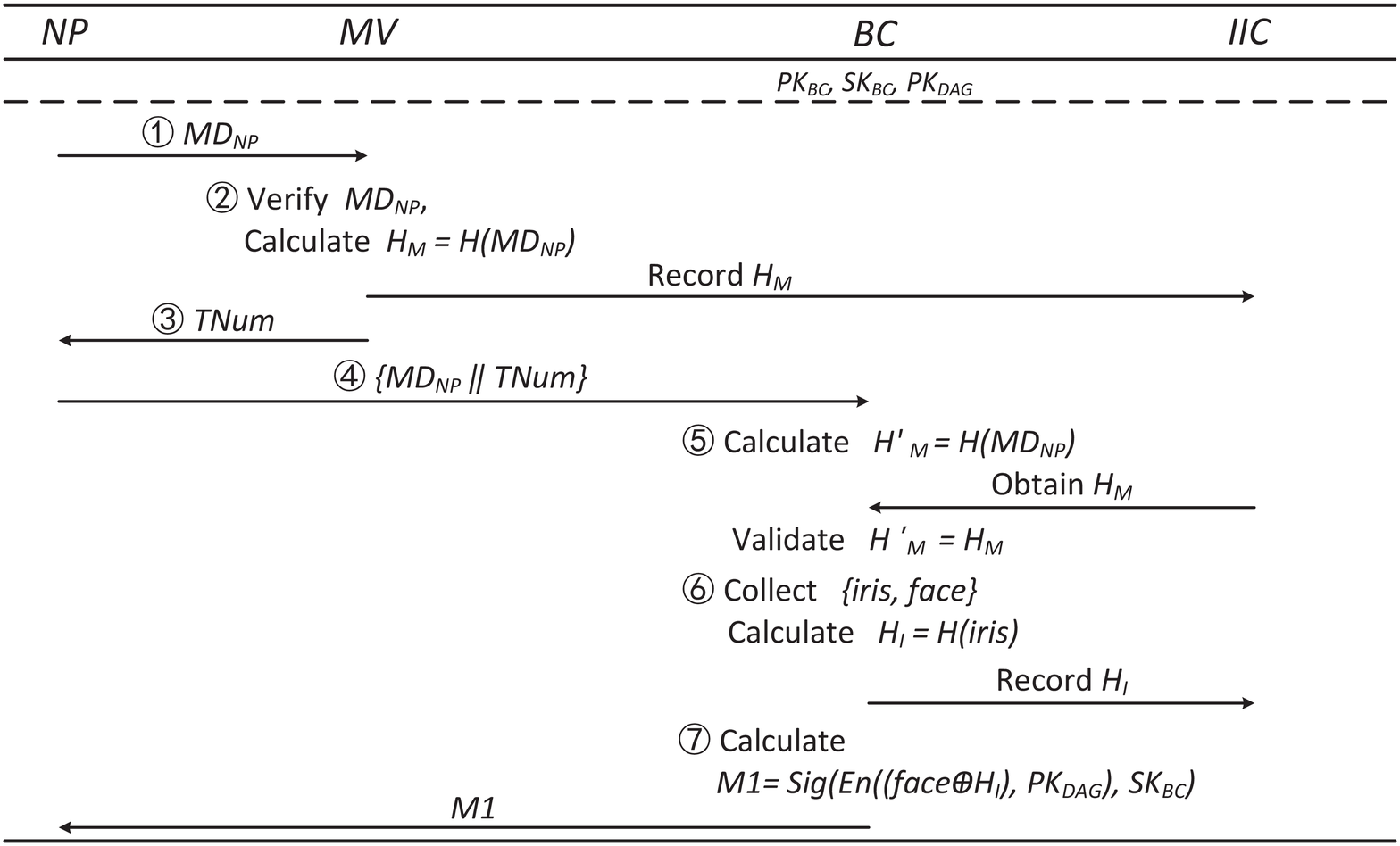}        
    \end{center}
    \caption{The flow chart of digitizing a NP\label{fig2}}
\end{figure}

To prepare for the generation of a DA, the NP sends the metadata to the ICVG for digitizing himself/herself. 
Here, the MV verifies the metadata and records the proof of metadata to the IIC. 
While the BC collects the NP's biometric data, writes the proof of the biometric data to the IIC and sends the NP a digitization credential. 
The whole process is shown in Figure~\ref{fig2}.
\begin{figure*}[!htbp]
    \begin{center}
        \includegraphics[width=\linewidth]{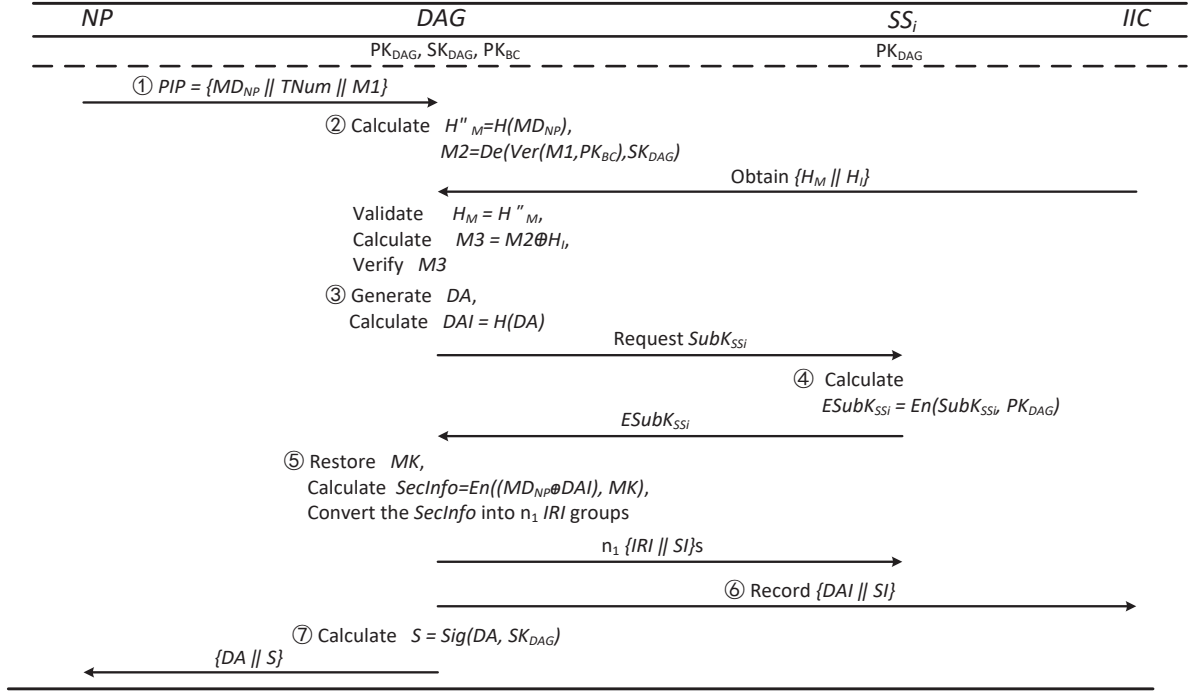}
    \end{center}
    \caption{The flow chart of generating a DA\label{fig3}}
\end{figure*}	

\textbf{STEP D1.} A NP presents the certificates, such as ID card, passport, etc., and provides the metadata ($MD_{NP}$, including name, id number, address and gender) to the MV face to face, as shown in step \ding{192}. 

\textbf{STEP D2.} The MV verifies the authenticity of the $MD_{NP}$ with the certificates. If it is confirmed, the MV calculates the proof $H_M = H(MD_{NP})$ and sends a metadata verification transaction (TM, as shown in the Equation (\ref{eq9})) to the IIC, as shown in step \ding{193}. 
\begin{equation}
    TM=(Tid, Tin[PK_{MV},\phi, \phi], Tout[PK_{BC}, H_M, \omega]) \label{eq9}
\end{equation}

In the Equation (\ref{eq9}), according to the paper \cite{lyu2020sbac}, Tid represents the transaction number of the TM. The input array Tin[] consists of three parts, the input address, the previous transaction and the input script. And the $PK_{MV}$,  the input address, is the initiator's public key, since TM is the original transaction, both the last transaction of the TM and the input script are denoted to the $\Phi$. The output array Tout[] is composed of three parts, where the $PK_{BC}$ is the accepter's address, $H_M$ is the data to be recorded in the IIC and $\omega$ is an out-script used to signature the TM. 

\textbf{STEP D3.} The MV sends the transaction number (TNum) of TM to the NP, as shown in step \ding{194}. 

\textbf{STEP D4.} The NP sends the $MD_{NP}$, TNum to the BC face to face for authentication, as shown in step \ding{195}. 

\textbf{STEP D5.} The BC calculates $H_M' = H(MD_{NP})$, uses TNum to find the $H_M$ of TM recorded in the IIC and checks whether $H_M'=H_M$ is satisfied. If not, the BC aborts and it is shown in step \ding{196}. 

\textbf{STEP D6.} The BC collects two kinds of biological characteristics, where the one is as a permanent proof of NP's existence in cyberspace and the other one is used to activating the DA. Specifically, the permanent one should be unbreakable and needs not to be collected frequently, therefore, we choose the iris data. While for frequently using the DA to access network services, an easy-to-collect face biometric is introduced. 
And then, the BC calculates the $H_I = H(iris)$ and sends a iris verification transaction (TI, as shown in (\ref{eq10})) to the IIC, as shown in step \ding{197}. 
\begin{equation}
    TI=(Tid, Tin[PK_{BC}, TM, \varphi], Tout[PK_{DAG}, H_I, \omega]) \label{eq10}
\end{equation}	

In the Equation (\ref{eq10}), the Tid is the transaction number of the TI, the $PK_{BC}$ is the creater's address of the TI, the TM is the previous transaction, the $PK_{DAG}$ is the accepter's address of the TI, and the $H_I$ is the data to be recorded in the IIC. 

\textbf{STEP D7.} As the Equation (\ref{eq13}) shows, the BC encrypts the face data with the $PK_{DAG}$ and then signatures it with the $SK_{BC}$ to generate the $M1$. 
\begin{equation}
    M1 = Sig(En((face \oplus H_I),PK_{DAG}),SK_{BC})\label{eq13}
\end{equation}

Finally, the BC sends $M1$ to the NP, as shown in step \ding{198}. 
\subsection{Generation}\label{subsec4.3}
After the digitization, the DAGG can generate a DA for the NP and the process consists of seven steps. At first, the NP applies to the DAGG for a DA. 
Secondly, the DAG verifies the authenticity of NP's identity with the proof information in the IIC. 
Thirdly, the DAG generates a DA and requests the SSs' subkeys. 
Then, the SSs send the encrypted subkeys to the DAG. 
Next, the DAG restores the $MK$ to generate the $SecInfo$, and splits it into multiple backup information. 
Afterwards, the DAG records the proof of DA in the IIC, and lastly sends the DA to the NP, as shown in Figure~\ref{fig3}. 

\textbf{STEP G1.} The NP sends the physical identity proof $PIP = \{MD_{NP}, TNum, M1\}$ to the DAG, as shown in step \ding{192}. 

\textbf{STEP G2.} The DAG calculates $H_M''=H(MD_{NP})$ and M2 by Equation (\ref{eq21}). 
\begin{equation}
    M2 = De(Ver(M1, PK_{BC}), SK_{DAG})\label{eq21}
\end{equation}

Afterwards, the DAG obtains the $H_M$ and the $H_I$ with TNum from the IIC and checks whether $H_M'' = H_M$ is met. 
At last, the DAG calculates $M3= M2 \oplus H_I$, and verifies the living face biometric of NP with the $M3$, as shown in step \ding{193}. 

\textbf{STEP G3.} If the NP's identity is confirmed, according to the Algorithm \ref{a1}, the DAG selects corresponding code modules (dynamic verification, file transfer, etc.) from the code library to get the DA with the digital avatar-i seed ($DAS$) which is produced from $M3$ by the algorithm in the paper of \cite{juels2006fuzzy}. 
The DA is divided into $k$ modules and the selected code modules are combined together in order. 
Then, the DAG calculates the identifier $DAI = H(DA)$ and requests all SSs to send their respective $SubK_{SS}$, as shown in step \ding{194}.
\begin{algorithm}
    \setstretch{1.2}
    \caption{Digital Avatar-i Generation}\label{a1}
    \begin{algorithmic}[1]
        \REQUIRE \

        The digital avatar-i seed $DAS$\

        The code module template $CMT_i$, $1 \leq i \leq k$
        \ENSURE \

        The digital avatar-i $DA$
        \STATE $len = strlen(DAS)/k$;
        \STATE $DA = 0$, $m = 0$, $n = len$;	
        \FOR{$i=0; i \leq k-1;i++$}
        \STATE $str=DAS.substring(m+len\times i,n+len\times i)$;
        \STATE  $a[i]=Decimal(str)\ mod\ (num[i])$;\

			//Decimal is a function that converts a string to 
					
			//a decimal

			//num[i] is the number of code module templates 

			//available in $CMT_i$
        \STATE $DA=Combine(DA, {CMT_i(a[i])});$\

			//Combine is a function that splices code 

			//modules in order. 
        \ENDFOR
        \RETURN $DA;$

    \end{algorithmic}
\end{algorithm}

\textbf{STEP G4.} Each $SS_i$ encrypts subkey to get $ESubK_{SS_{i}}$ by the Equation (\ref{eq22}) and sends it to the DAG, as shown in step \ding{195}. 
\begin{equation}
	ESubK_{SS_{i}} = En(SubK_{SSi}, PK_{DAG})\label{eq22}
\end{equation}

\textbf{STEP G5.} The DAG calculates at least $t_1-1$ $SubK_{SS_{i}} = De(ESubK_{SS_{i}}, SK_{DAG}),i=1,2,...,t_1-1$ and constructs the Lagrangian interpolation formula (as shown in the Equation (\ref{eq23})) with these $SubK_{SS_{i}}$s to restore the $MK=f'(0)$. 
\begin{equation}
	f'(x)=\sum_{i=1}^{t_{1}}y_i\prod_{j=1,j \neq i}^{t_{1}}\frac{(x-x_j)}{x_i-x_j}\label{eq23}
\end{equation}	

Afterwards, the DAG generates the $SecInfo = En((MD_{NP} \oplus DAI), MK)$ and expands SecInfo to $n\times t_{1}\times b$ bytes by filling high bits with zero. 
The DAG constructs $n$ polynomials, as showed in the Equation (\ref{eq24}). 

\begin{equation}
	\left\{
		\begin{aligned}
			&F_1(x)=m_0+m_1x+\cdots+m_{t_{1}-1}x^{t_{1}-1}mod(N)\\
			&F_2(x)=m_{t_1}+m_{t_1+1}x+\cdots+m_{2t_{1}-1}x^{t_{1}-1}mod(N)\\
			&\vdots\\ 
			&\makecell[r]{F_n(x)=m_{(n-1)t_{1}}+m_{(n-1)t_{1}+1}x+\cdots+m_{nt_{1}-1}x^{t_1-1}\\mod(N)}\label{eq24}
		\end{aligned}            
	\right.   
\end{equation}	

In the Equation (\ref{eq24}), the SecInfo is divided into $n\times t_{1}$  coefficients in order and each coefficient is $b$ bytes. The $N$ is a prime number bigger than any coefficient $m_i$, and the length of $N$ is $b+1$ bytes. 

\begin{figure}[!htbp]
	\begin{center}
		\includegraphics[scale=0.45]{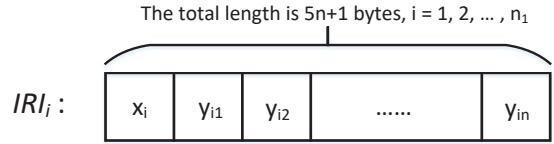}            
	\end{center}
	\caption{The format of the $IRI_{i}$\label{fig4}}
\end{figure}

The DAG substitutes $n_{1}\ x$s into each polynomial in the Equation (\ref{eq24}) to calculate $n\times n_{1}$ points $(x_i, y_{ij}), i=1,2,...n_1, j=1,2,...,n$ 
(since $x_i$ performs multiple exponentiation operations, the length of $x_i$ is set to one byte to reduce computational overhead, while the length of $y_{ij}$ is set to five bytes to avoid collisions between these points. If the lengths of $x_i$ and $y_{ij}$ are too short, the high bits are filled with zero). 
Further, the DAG divides all the points into $n_1$ groups and $n$ points in each group come from the different $F(x)$s. 
It is worth mentioning that the $x_i$ in these $n$ points are the same, and the $n$ points $(x_i, y_{ij})$ are combined to form a set of identity restoration information (IRI), which is as shown in the Figure~\ref{fig4}. 
At last, the DAG transmits $n_1$ sets of IRI and storage index (SI) to all the SSs, as shown in step \ding{196}. 

\textbf{STEP G6.} If the $SS_i$ receives the IRI and SI, he/she sends a response to the DAG. 
When DAG confirms that more than half of SSs have received IRI and SI, he/she writes a DA generation transaction (TDA, as shown in (\ref{eq20})) in the IIC, as shown in step \ding{197}. 
\begin{equation}
	\begin{split}
		TDA=(Tid, Tin[PK_{DAG}, TI, \varphi], Tout[PK_{DAG}, \\
		(DAI \parallel  SI), \omega]) \label{eq20}
	\end{split}
\end{equation}

In the Equation (\ref{eq20}), the Tid is the transaction number of the TDA, the TI is the previous transaction, the $PK_{DAG}$ is the creater and the accepters' address of the TDA, and the $(DAI \parallel SI)$ is the data to be recorded in the IIC. 

\textbf{STEP G7.} The DAG calculates the DA's proof $S=Sig(DA,SK_{DAG})$ and sends it with the DA to the NP, as shown in step \ding{198}. 
\subsection{Interaction}\label{subsec4.4}
After receiving the DA, the NP can access various services provided by Dapps built on DABC through it. At first, the NP activates the DA through live face recognition. And then, the DAI or a random string can be selected by the activated DA as the identifier for the NP to participate in activities in cyberspace. It is worth mentioning that all behaviors of the NP accessing network services will be recorded in the DABC for future audit. 

In a word, the main work of this phase is to use DA for authentication and authorization, which requires unlinkable identity, informed consent and the right to be forgotten, etc. 
However, limited by space, details such as the protocol process, algorithms and data format will be given in our future work. 
\subsection{Accountability}\label{subsec4.5}
\begin{figure}[!htbp]
	\begin{center}
		\includegraphics[width=1.0\columnwidth]{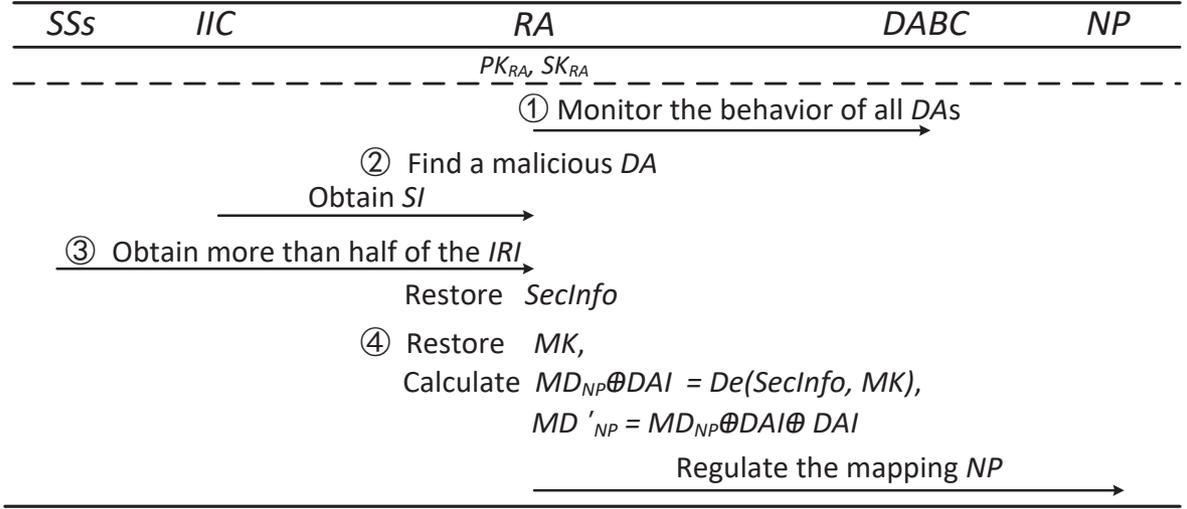}            
	\end{center}
	\caption{The flow chart of regulating the NP\label{fig5}}
\end{figure}

When a malicious behavior of a DA occurs, the RAG can discover the mapping NP by inquiring the IIC and finding the metadata of the NP in the DAGG with the joint participation of multiple RAs. 
Then, the RAG can regulate all the historical behaviors of the malicious NP, as shown in Figure~\ref{fig5}. 

\textbf{STEP S1.} All the RAs are monitoring the DAs' behavior in DABCs, as shown in step \ding{192}. 

\textbf{STEP S2.} When the $RA_i$ finds a suspicious behavior of a DA, he/she can start the accountability mechanism. 
Firstly, $RA_i$ inquires SI from the IIC with DAI and writes a audit transaction (TA, as shown in (\ref{eq25})) in the IIC, as shown in step \ding{193}. 
\begin{equation}
	\begin{split}
		TA=(Tid, Tin[PK_{RA_{i}}, TA, \phi], Tout[PK_{RA_i}, \\
			(timestamp \parallel DAI), \omega])\label{eq25}
	\end{split}
\end{equation}

In the Equation (\ref{eq25}), the Tid is the transaction number of the TA, the TA in the Tin[] is the previous audit transaction, the $PK_{RA_{i}}$ is the creater and the accepters' address of this TA, and the $(timestamp \parallel DAI)$ is the data to be recorded in the IIC. 

\textbf{STEP S3.} Then, the $RA_i$ finds at least $t_1$ IRIs stored in SSs with the SI. 
And all the obtained IRIs are processed as follows: 
\ding{192}The $RA_i$ decomposes each $IRI_j, j=1,2,..., t_1$ into $n$ points, where the structure of the $IRI_j$ is shown in the Figure~\ref{fig4} and the decomposed point set is shown in the Equation (\ref{eq30}); 
	\begin{equation}
		\left\{
			\begin{array}{cccc}
				(x_1,y_{1,1})   & (x_1,y_{1,2})   &  \cdots   & (x_1,y_{1,n})   \\
				(x_2,y_{2,1})   & (x_2,y_{2,2})   &  \cdots   & (x_2,y_{2,n})   \\
				\vdots          & \vdots          &  \ddots   & \vdots          \\
				(x_{t_1},y_{t_1,1}) & (x_{t_1},y_{t_1,2}) & \cdots  & (x_{t_1},y_{t_1,n})\label{eq30}
			\end{array}
			\right\}
	\end{equation}

    \ding{193}Then substituting the $t_1$ points $(x_1, y_{1,1}),(x_2,y_{2,1}),$\\$...,$ $(x_{t_1},y_{t_{1},1})$ into the Lagrangian interpolation formula (\ref{eq26}) to obtain the polynomial $F_1(x)$; 
    \begin{equation}
        F'(x)=\sum_{i=1}^{t_{1}}y_i\prod_{j=1,j \neq i}^{t_{1}}\frac{(x-x_j)}{x_i-x_j}\label{eq26}
    \end{equation}	

    \ding{194}Similar to the step \ding{193}, the $RA_i$ obtains the polynomials $F_2(x)$, $\cdots$, $F_n(x)$. 

    After the $n$ polynomials are obtained, the $SecInfo$ is restored by splicing the coefficients of these polynomials in order, as shown in step \ding{194}. 

	\textbf{STEP S4.} The $RA_i$ initializes an audit request to all other $RA_j$s. Each $RA_j$ encrypts his/her $SubK_{RA_{j}}$ by the Equation (\ref{eq27}) and sends the 
    $ESubK_{RAj}$ to the $RA_i$. 
    \begin{equation}
        \begin{aligned}
            ESubK_{RA_{j}}=En(SubK_{RA_{j}}, PK_{RA_{i}})\\
            (j=1,2,...,n_2;j \neq i)\label{eq27}            
        \end{aligned}
    \end{equation}		

    When more than $t_2$ $RA_j$s respond, the $RA_i$ decrypts the $ESubKey_{RA}$ one by one using the Equation (\ref{eq28}). 
    \begin{equation}
        \begin{aligned}
            SubK_{RA_{j}}=De(ESubK_{RA_{j}}, SK_{RA_{i}})\\
            (j=1,2,...,t_2;j \neq i)\label{eq28}            
        \end{aligned} 
    \end{equation}	

	Then, the $RA_i$ constructs the Lagrangian interpolation formula with the $t_2$ $SubK_{RA}$s, as shown in the Equation (\ref{eq29}), to calculate the $MK=f'(0)$. 
    \begin{equation}
        f'(x)=\sum_{i=1}^{\lfloor t_{2} \rfloor}y_i\prod_{j=1,j \neq i}^{\lfloor t_{2} \rfloor}\frac{(x-x_j)}{x_i-x_j}\label{eq29}
    \end{equation}	

	After that, the $RA_i$ decrypts the $SecInfo$ to get the $(MD_{NP}\oplus DAI) = De(SecInfo, MK)$, and gets $MD_{NP}' = (MD_{NP}\oplus DAI) \oplus DAI$. 
    So far, the $RA_i$ can discover the malicious NP and regulate him/her through the historical behaviors in the DABC, as shown in step \ding{195}. 
\section{Security analysis}\label{sec5}
In this section, we discuss the security of the proposed scheme.
\subsection{Conditional anonymity}\label{subsec5.1}
In this scheme, the metadata of a NP ($MD_{NP}$) is hidden in the $SecInfo$ by the $DAI$ and $MK$, and the other entities except RA cannot recover it without the $DAI$ and $MK$. The $DAI$ is recorded on the IIC, while the shamir (t, n) threshold algorithm protects the $MK$. Therefore, the scheme realizes the anonymity of entities other than the RA. On the other hand, we allow at least $t$ RAs to restore the $MK$ for revealing the $MD_{NP}$ by the Lagrangian interpolation formula. In short, the conditional anonymity is achieved in our scheme. 
\subsection{Anti-sybil attack}\label{subsec5.2}
In this scheme, each NP needs to digitize himself/herself through the ICVG before applying for a DA, where the authenticity of the NP's $MD_{NP}$ is verified by the MV with NP's certificates and the NP's biometric data is collected by the BC as a proof of unique identity. In addtion, the hash of the $MD_{NP}$ and biometric data (iris) are permanently recorded on the IIC. In this way, it can be ensured that each NP has one and only DA and the sybil attack is avoided. 
\subsection{Tamper-proof}\label{subsec5.3}
During the digitization phase, the NP is required to provided $MD_{NP}$ face-to-face, therefore, the tampered $MD_{NP}$ submitted by the adversary cannot be verified by the MV with the NP's certificates. And the consortium blockchain records the proof of $MD_{NP}$ and biometric data, no one can easily erase the NP's information. In addition, the DAG calculates the signature $S=Sig(DA,SK_{DAG})$ to prevent the adversary to tamper with the DA. 
\subsection{Non-repudiation}\label{subsec5.4}
For each DA, the RAG can find the corresponding $SecInfo$ through the data in the IIC and SSs. Then, the RAG can calculate the $MK$ with the participation of multiple RAs, and get $MD_{NP}\oplus DAI= De(SecInfo,MK)$. Using the known $DAI$ in the IIC, the RAG can obtain the $MD_{NP}= MD_{NP}\oplus DAI \oplus DAI$, and track the mapping NP with it. That is, a NP cannot deny his/her malicious historical behavior. 
\subsection{Impersonation-resistance}\label{subsec5.5}
When accessing a DA, the NP's biometric data needs to be verified by the DA in advance. In addition, the DA includes a dynamic verification module which will issue dynamic verification requests to the NP from time to time. Once the NP fails to pass the verification, the DA will be locked. Therefore, even if an adversary obtains a DA that does not belong to him/her, he/she cannot use it to participate in network activities. 
\subsection{Data security}\label{subsec5.6}
The metadata of a NP ($MD_{NP}$) and the hash of a DA ($DAI$) are firstly encrypted as the $SecInfo$, then divided into $n\times t_1$ parts, and finally converted into $n\times n_1$ points by shamir (t,n) threshold algorithm. In this way, the cost of obtaining a NP's information by an adversary is greatly increased. Further, the information transmitted between different entities, such as subkeys, are protected by asymmetric encryption algorithm. Therefore, the scheme guarantees the security of the data. 
\section{Performance analysis}\label{sec6}
This section analyzes the cost of our proposed NSSIA, and compares it with the above schemes \cite{takemiya2018sora,lee2019sims,zheng2019blockchain,bandara2021blockchain,maram2021candid} in terms of the SSI property in identity generation, computation cost, storage cost, and blockchain Gas cost. 
\subsection{Property analysis in identity generation}\label{subsec6.1}
Ten principles are proposed by Christopher Allen \cite{noauthor_path_nodate} to define an SSI model, which are Existence, Control, Access, Transparency, Persistence, Portability, Interoperability, Consent, Minimalization and Protection. 
It can be said that Allen's insights on SSI lays the foundation for the research of later generations. 

As the shortcomings of the centralized identity model have been revealed in recent years, more and more scholars have invested in the research of SSI, and their work can be seen from literatures \cite{ferdous2019search,muhle2018survey}. 
It is worth mentioning that before this article is written, Ferdous et al. \cite{ferdous2019search} had introduced in detail the insights of various scholars on SSI. 
At the same time, they put forward their own views on the properties of SSI. 
They divided self-sovereign identity into five categories, with a total of seventeen properties. M{\"u}hle et al. \cite{muhle2018survey} analyzed the work of Christopher Allen, and then studied four basic components for having a deeper understanding of the concept of SSI. 

\begin{figure}[!htbp]
  \begin{center}
      \includegraphics[width=1.0\columnwidth]{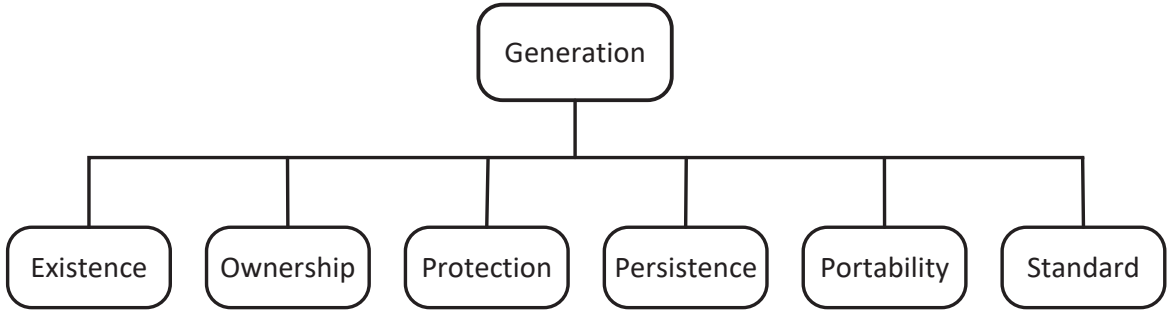}      
  \end{center}
  \caption{Taxonomy of the Identity Generation\label{fig6}}
\end{figure}

As the identity generation is the critical step for users to enter cyberspace, we focus on analyzing the properties that SSI needs to follow at this stage, as shown in Figure~\ref{fig6}. 
And these properties are depicted next. 

\begin{enumerate}
    \item[$\bullet$] \textbf{Existence.} Digital identities should be strictly verified before registration to ensure that each digital identity has a corresponding physical entity. 
    \item[$\bullet$] \textbf{Ownership.} Digital identities can only be held and controlled by users. 
    \item[$\bullet$] \textbf{Protection.} The registration of digital identities should pay attention to protecting user privacy and avoiding the identity link between the physical world and the cyberspace. At the same time, the design of SSI model should prevent fraudulent use of identity by others.          
    \item[$\bullet$] \textbf{Persistence.} The digital identity should exist forever if the user does not take the initiative to revoke. 
    \item[$\bullet$] \textbf{Portability.} When the user's device is replaced or the system's infrastructure is updated, the user's data can be easily transferred to the new device. 
    \item[$\bullet$] \textbf{Standard.} The SSI model should comply with the laws and regulations of various countries and international standards, such as GDPR, DID, etc. 
\end{enumerate}

Next, we compare our scheme with the previously mentioned schemes on these properties, as shown in Table~\ref{tab2}. 

\begin{table*}[!htbp]
    \centering
    \caption{Property comparison\label{tab2}}
    \renewcommand\tabcolsep{10.0pt}
    \begin{threeparttable}
		\def\arraystretch{1.15}
        \begin{tabular}{ccccccc}
            \toprule[1.5pt]
            Schemes   & Existence & Ownership & Protection & Persistence & Portability  & Standard \\
            \midrule[0.8pt]
            \cite{takemiya2018sora}      & N & Y  & P & Y                         & P                         & P \\
            \cite{lee2019sims}           & N & Y  & P & Y                         & \rule[3pt]{0.3cm}{0.1em}  & P \\
            \cite{zheng2019blockchain}   & Y & P  & Y & \rule[3pt]{0.3cm}{0.1em}  & \rule[3pt]{0.3cm}{0.1em}  & N \\ 
            \cite{bandara2021blockchain} & Y & Y  & P & Y                         & P                         & P \\
            \cite{maram2021candid}       & P & Y  & P & Y                         & \rule[3pt]{0.3cm}{0.1em}  & P \\
            ours                         & Y & Y  & Y & Y                         & Y                         & Y \\
            \bottomrule[1.5pt]	
        \end{tabular}
        \begin{tablenotes}
            \footnotesize
            \item[a] The "Y" and "N" symbols respectively indicate that a certain property is satisfied or not in the corresponding scheme. The "P" symbol indicates that part of the property is satisfied. The "\rule[3pt]{0.3cm}{0.1em}" symbol shows that there is no obvious information about whether the property is satisfied or not.  
        \end{tablenotes}	
    \end{threeparttable}
  \end{table*}

From Table~\ref{tab2}, the properties of "Ownership" and "Persistence" are satisfied in \cite{takemiya2018sora}, and the part of "Protection", "Portability" and "Standard" properties are met, however, the "Existence" property is unsatisfied. 
In \cite{lee2019sims}, "Ownership" and "Persistence" properties, as well as the part of "Protection" and  "Standard" properties are fulfilled, while the "Existence" property is unsatisfied and the rest is uncertain. 
In \cite{zheng2019blockchain}, "Existence" and "Protection" properties as well as the part of "Ownership" property is satisfied, but the "Standard" property is unsatisfied and the others is doubtful. 
The properties of "Existence", "Ownership" and "Persistence" properties are satisfied in \cite{bandara2021blockchain}, and the part of "Protection", "Portability" and "Standard" is met. 
Maram et al. \cite{maram2021candid} fulfills the properties of "Ownership" and "Persistence", as well as the part of "Existence", "Protection" and "Standard" properties, but the rest is doubtful. 
Compared with them, these properties are all realized in our scheme. 
\subsection{Computation Cost}\label{subsec6.2}
\subsubsection{The NSSIA}\label{subsubsec6.2.1}
Our scheme includes five phases, namely initialization, digitization, generation, interaction and accountability. Since the initialization phase is executed only once and the interaction phase is not the point, we do not evaluate these two phase, and we mainly focus on the other three phases. Our scheme is run on a PC with windows 10, Intel(R) Core(TM) i5-1035G1 CPU @ 1.00GHz and RAM 16G. We use Java 8.0 and Python 3.9 to evaluate the computation cost, where we choose the SHA-1 algorithm, the AES-128 algorithm, and the Secp256k1 elliptic curve. In order to balance security and computational cost, the $n_1$ and $n_2$ which is the number of SSs and RAs respectively are set to 5 in this simulation, and the corresponding $t_1$ and $t_2$ are 3. 
In addition, the length of the $SubK_{e}$ is $17$ B. The $MD_{NP}$, $DA$, $n$ and $b$ are $256$ B, $1$ KB, 20 and $5$ B separately. That is, the length of $SecInfo$ is $t_1\times n \times b=300$ B. According to the data in the paper \cite{bobkowska2019incorporating}, the length of iris and face is 25 KB and 30 KB respectively. 

To elaborate the computation cost clearly, a series of computational notations are defined: 
\begin{itemize}
	\item $T_{H1}$ denotes the SHA-1 operation. $T_{H1_1}$, $T_{H1_2}$, and $T_{H1_3}$ are the computation cost of performing the SHA-1 operation when the parameter sizes are $256$ B, $1$ KB, and $25$ KB, respectively, and they are $0.0169$ ms, $0.0353$ ms, $0.0748$ ms. 
	\item $T_{S1}$ indicates the AES-128 encryption/ decryption algorithm and the computation cost is $0.4186$ ms when the parameter size is $256$ B. 
	\item $T_{AE}$ represents the ECC encryption algorithm. $T_{AE_1}$ and $T_{AE_2}$ are the computation cost of performing the ECC encryption algorithm when the parameter sizes are $17$ B and $30$ KB, respectively, and they are $0.0015$ ms, $0.0024$ ms. 
	\item $T_{AD}$ expresses the ECC decryption algorithm. $T_{AD_1}$ and $T_{AD_2}$ are the computation cost of performing the ECC description algorithm when the parameter sizes are $17$ B and $30$ KB, respectively, and they are $0.034$ ms, $0.0401$ ms. 
	\item $T_{Sig}$ serves as the ECDSA signature algorithm. $T_{Sig_1}$ and $T_{Sig_2}$ are the computation cost of performing the ECDSA signature algorithm when the parameter sizes are $1$ KB and $30$ KB, respectively, and they are $0.0014$ ms, $0.0283$ ms. 
	\item $T_{Ver}$ is the ECDSA verification algorithm. $T_{Ver_1}$ and $T_{Ver_2}$ are the computation cost of performing the ECDSA verification algorithm when the parameter sizes are $1$ KB and $30$ KB, respectively, and they are $0.0007$ ms, $0.001$ ms. 
	\item $T_L$ is the Lagrangian interpolation algorithm and the computation cost is $0.0101$ ms when the threshold value is $t=3$. 
\end{itemize}

Since the time cost of XOR operation, split operation and splicing operation is negligible, it is not taken into consideration. 

The computation cost of different phases in our scheme is shown in the Table~\ref{tab3}. 
\begin{table}[!htbp]
    \centering
    \caption{Computation cost on different phases\label{tab3}}
    \renewcommand\tabcolsep{8.0pt}
        \scalebox{1}{
		\def\arraystretch{1.15}
        \begin{tabular}{cc}
            \toprule[1.5pt]
            Phases &  Computation Cost (ms) \\
            \midrule[0.8pt]
            Digitization & \makecell[c]{$2T_{{H1}_1}+T_{{H1}_3}+T_{Sig_2}$\\$+T_{AE_2}=0.1393$}    \\
            \midrule[0.5pt]
            Generation   & \makecell[c]{$T_{{H1}_1}+T_{{H1}_2}+2T_{AE_1}$ \\$+T_{Ver_2}+T_{Sig_1}+T_L+T_{S1}$\\$+2T_{AD_1}+T_{AD_2}=0.5944$}     \\
            \midrule[0.5pt]
            Accountability   &   \makecell[c]{$2T_{AE_1}+2T_{AD_1}+21T_L$\\$+T_{S1}=0.7017$}  \\
            \bottomrule[1.5pt]
        \end{tabular}}	
\end{table}

In the Table~\ref{tab3}, three SHA-1 operations, one ECDSA signature algorithm and one ECC encryption algorithm are performed in the Digitization phase and the time cost is $2T_{{H1}_1}+T_{{H1}_3}+T_{Sig_2}+T_{AE_2}=0.1393$ ms. Two SHA-1 operations, two ECC encryption algorithms, one ECDSA verification algorithm, one ECDSA signature algorithm, one Lagrangian interpolation algorithm, one AES encryption and three ECC decryption algorithms are performed in the Generation phase and the time cost is $T_{{H1}_1}+T_{{H1}_2}+2T_{AE_1}+T_{Ver}+T_{Sig_1}+T_L+T_{S1}+2T_{AD_1}+T_{AD_2}=0.5944$ ms. The Accountability phase consists of two ECC encryption algorithms, two ECC decryption algorithms, twenty-one Lagrangian interpolation algorithms and one AES decryption algorithm and the time cost is $2T_{AE_1}+2T_{AD_1}+21T_L+T_{S1}=0.7017$ ms. To generate a unique DA for each NP, the total $0.1393+0.5944+0.7017=1.4354$ ms is used. 
\subsubsection{Computation cost comparison}\label{subsubsec6.2.2}
We compare with other schemes in terms of the generation phase and the accountability phase, as shown in Table~\ref{tab4}. 
\begin{table*}[!htbp]
	\centering
	\caption{Computation cost comparison\label{tab4}}
	\renewcommand\tabcolsep{6pt}
	\begin{threeparttable}
		\scalebox{1}{
			\def\arraystretch{1.15}
			\begin{tabular}{ccc}
				\toprule[1.5pt]
				Scheme &  Generation Cost (ms) & Accountability Cost (ms)\\
				\midrule[0.8pt]
				{\cite{takemiya2018sora}} & \makecell[c]{$T_P+T_{H2}+T_{S2_1}+T_{S2_2}=40.5941$} & \rule[3pt]{0.3cm}{0.1em}\\
				\midrule[0.5pt]
				{\cite{lee2019sims}} & \makecell[c]{$T_{{H1}_1}+T_{Sig_1}+T_Z=18217.9183$} & \rule[3pt]{0.3cm}{0.1em}\\
				\midrule[0.5pt]
				{\cite{zheng2019blockchain}} & \makecell[c]{$3T_{AE_2}+3T_{Sig_1}+3T_{Ver_1}+2T_{S1}+3T_{AD_2}=0.971$} & \rule[3pt]{0.3cm}{0.1em} \\
				\midrule[0.5pt]
				{\cite{bandara2021blockchain}} & $T_B+2T_R=6.0892$ & \rule[3pt]{0.3cm}{0.1em} \\
				\midrule[0.5pt]
				{\cite{maram2021candid}} & \makecell[c]{$T_O+T_{ZKP}=2350$} & \makecell[c]{$T_M=1501540$} \\
				\midrule[0.5pt]
				Ours & \makecell[c]{$T_{{H1}_1}+T_{{H1}_2}+2T_{AE_1}+T_L+T_{Sig_1}+T_{Ver_2}+2T_{AD_1}$ \\$+T_{S1}+T_{AD_2}=0.5944$} &\makecell[c]{$2T_{AE_1}+2T_{AD_1}+21T_L+T_{S1}=0.7017$} \\
				\bottomrule[1.5pt]
			\end{tabular}
		}
		\begin{tablenotes}
			\footnotesize
			\item[a] \bm{$T_P$} is the PBKDF2 algorithm and the computation cost is $40.18$ ms when the parameter size is $8$ B and the number of rounds is $61337$; \bm{$T_{H2}$} indicates the SHA-256 operation, and the computation cost is $0.035$ ms when the parameter size is $32$ B; \bm{$T_{S2}$} denotes the AES-256 encryption/ decryption algorithm. \bm{$T_{S2_1}$} and \bm{$T_{S2_2}$} are the computation cost of performing the AES-256 encryption/ decryption algorithm when the parameter sizes are $64$ B and $128$ B, respectively, and they are $0.187$ ms, $0.1921$ ms. 
			\item[b] \bm{$T_Z$} denotes the time complexity of zero-knowledge proof. According to the definition of identity information in our scheme, we refer to the performance simulation of \cite{lee2019sims}, and the value of \bm{$T_Z$} is $18271.9$ ms. 
			\item[c] \bm{$T_B$} serves as the base58 encoding algorithm and the computation cost is $0.086$ ms when the parameter size is $256$ B; \bm{$T_R$} expresses the RSA signature algorithm and the computation cost is $3.002$ ms when the parameter size is $450$ B. 
			\item[d] \bm{$T_O$}, \bm{$T_{ZKP}$} and \bm{$T_M$} indicates the time complexity of an oracle, zero-knowledge proof and secure multiparty computation respectively. According to the definition of identity information in our scheme, we refre to the performance simulation of \cite{maram2021candid}, and the value of \bm{$T_O$}, \bm{$T_{ZKP}$} and \bm{$T_M$} are $1400$ ms, $950$ ms and $1501540$ ms respectively. 
		\end{tablenotes} 
	\end{threeparttable}
  \end{table*}

In Table~\ref{tab4}, one PBKDF2 algorithm, one SHA-256 operation and two AES-256 encryption algorithm are performed in generation phase in \cite{takemiya2018sora}, and the time cost is $T_P+T_{H2}+T_{S2_1}+T_{S2_2}=40.5941$ ms. 
In \cite{lee2019sims}, one SHA-1 operation, one ECDSA signature algorithm and one zk-SNARK algorithm are performed in generation phase, in which the time cost is $T_{{H1}_1}+T_{Sig_1}+T_Z=18217.9183$ ms. 
Zheng et al. \cite{zheng2019blockchain} performs two AES-128 encryption algorithms, three ECC encryption algorithms, three ECC decryption algorithms, three ECDSA signature algorithms and three ECDSA verification algorithms in generation phase, and the time cost is $3T_{AE_2}+3T_{Sig_1}+3T_{Ver_1}+2T_{S1}+3T_{AD_2}=0.971$ ms. 
In \cite{bandara2021blockchain}, one base58 encoding algorithm and two RSA signature algorithms are performed in generation phase, and the time cost is $T_B+2T_R=6.0892$ ms. 
One oracle operation and one ZKP operation are performed in generation phase in \cite{maram2021candid}, and the time cost is $T_O+T_{ZKP}=2350$ ms. 
As shown in Section~\ref{subsec6.1}, the computation cost is $T_{{H1}_1}+T_{{H1}_2}+2T_{AE_1}+T_L+T_{Sig_1}+T_{Ver_2}+2T_{AD_1}+T_{S1}+T_{AD_2}=0.5944$ms in generation phase in our scheme. 

For the accountability phase, since the corresponding mechanism has not been designed in the paper \cite{takemiya2018sora,lee2019sims,zheng2019blockchain,bandara2021blockchain}, the audit cost cannot be given. 
While in \cite{maram2021candid}, one secure multiparty computation operation is performed in accountability phase, where the audit cost is $T_M=1501540$ ms. 
As shown in Section~\ref{subsec6.1}, the accountability cost in our scheme is $2T_{AE_1}+2T_{AD_1}+21T_L+T_{S1}=0.7017$ ms. 
From Table~\ref{tab4}, our scheme has the lowest time overhead in both the generation phase and the accountability phase. 
\subsection{Storage cost}\label{subsec6.3}
We compare the storage cost with other schemes from the perspective of users, servers, and blockchain. The comparison result is shown in Table~\ref{tab5}. 
\begin{table}[!htbp]
    \centering
    \caption{Storage cost comparison\label{tab5}}
    \renewcommand\tabcolsep{1.5pt}
        \scalebox{1}{
		\def\arraystretch{1.15}
        \begin{tabular}{cccc}
            \toprule[1pt]
            Scheme & User(bytes) & Server(bytes) & Blockchain(bytes) \\
            \midrule[0.8pt]
            {\cite{takemiya2018sora}}  & $168$ & $108$ & $\textgreater 200$ \\
            {\cite{lee2019sims}}  & $16\sim32$ & $198$ & $102$\\
            {\cite{zheng2019blockchain}}  & $304$  & $\textgreater 10240$ & $20$\\
            {\cite{bandara2021blockchain}} & $\approx 10670$  & \rule[3pt]{0.3cm}{0.1em} & $800$ \\
            {\cite{maram2021candid}} & $\textgreater 150$ & \rule[3pt]{0.3cm}{0.1em} & \rule[3pt]{0.3cm}{0.1em} \\
            Ours & 0 & $505$ & $94$\\
            \bottomrule[1.5pt]
        \end{tabular}}
\end{table}

As shown in the Table~\ref{tab5}, users in \cite{takemiya2018sora} need to store a master key and a corresponding derived key, as well as the encrypted personal identity information, which are $168$ bytes. 
And users in \cite{lee2019sims} need to store a random value key for hash algorithm, and the storage cost is estimated to be $16-32$ bytes. 
In \cite{zheng2019blockchain}, three pairs of public and private keys, as well as a password used for symmetric encryption algorithm are stored locally by the user, which are $304$ bytes. 
And users in \cite{bandara2021blockchain} need to store a private key for signature algorithm and personal information (name, DID, photo, etc.) locally, which are exceeds $10670$ bytes. 
In \cite{maram2021candid}, a credential containing user's information is stored locally, and the storage cost is estimated to be over $150$ bytes. 
While with the help of the biometrics, there is no data such as keys need to be stored locally by users in our scheme. 

For a server, an estimated cost cannot be given in the paper \cite{bandara2021blockchain,maram2021candid}, because there is no detailed description. 
A pair of public and private keys encrypted by the AES-256 algorithm needs to be stored in the server in \cite{takemiya2018sora}, and the storage cost is $108$ bytes. 
In \cite{lee2019sims}, the user's identity information and related certificates are stored in the server, which are totally $198$ bytes. 
And the server in \cite{zheng2019blockchain} needs to store the user's information and the certificate containing the user's phone number, photo and so on, and the storage cost is exceeds $10240$ bytes. 
While the storage cost is $505$ bytes composed of $IRIs$ in our scheme. 
Because we divide the encrypted user information into multiple pieces based on the shamir(t, n) threshold algorithm and store them in different servers, so as to audit malicious users without revealing user privacy. 

The last is the storage cost of each scheme on the blockchain. 
Since there is no evidence that a blockchain is deployed in \cite{maram2021candid}, an estimated overhead cannot be given. 
In \cite{takemiya2018sora}, a public part of the claim used to proof the identity is written into the blockchain and the cost is exceeds $200$ bytes. 
In \cite{lee2019sims}, the hash value of the user's identity information and the corresponding certificate are recorded in the blockchain, which is $20+82=102$ bytes. 
And in \cite{zheng2019blockchain}, the hash value of the user's identity information is written into the blockchain and the storage cost is $20$ bytes. 
A DID proof with DID, name, signature, etc. is recorded in the blockchain in \cite{bandara2021blockchain}, which is $800$ bytes. 
In our scheme, the hash value of user's metadata, biometrics and digital identity, as well as the timestamp are recorded on the blockchain, which is $20+20+20+20+14=94$ bytes. 
From Table~\ref{tab5}, we achieve lower storage cost in the blockchain compared to the schemes \cite{takemiya2018sora,lee2019sims,bandara2021blockchain}. 
Although Zheng et al. \cite{zheng2019blockchain} has lower overhead than us, the data recorded on the blockchain in our scheme more intuitively shows the entire process of identity generation and accountability. 
In addition, the data is written to the blockchain by different entities, which decentralizes power of regulatory authorities and reduces the risk of information leakage compared to \cite{zheng2019blockchain}. 
In short, we liberate users in terms of storage, while servers and the blockchain have the necessary storage requirements to balance privacy and accountability, which are low enough for practical scenarios. 
\subsection{Blockchain Gas Cost}\label{subsec6.4}
Considering Gas cost as an important aspect to measure performance, we conduct detailed experiments in this regard. 
To visualize the execution cost of our smart contract, we evaluate its practical performance on a public Ethereum testnet (Rinkeby). 
We used the plugin Metamask in Chrome v100.0 explorer to access the Rinkeby testnet and the Remix, a browser-based IDE, to compile and deploy our smart contract. 
Rinkeby is built in April 2017 by the Ethereum Foundation and it uses the proof-of-authority consensus mechanism. 
Since the ether supply is controlled by several trusted parties and only they can write transactions on the blockchain, it can be considered a consortium blockchain. 
Hence, the waiting time for a transaction to be confirmed is relatively short to be ignored. 

\begin{figure}[!htbp]
    \begin{center}
        \includegraphics[width=1.0\columnwidth]{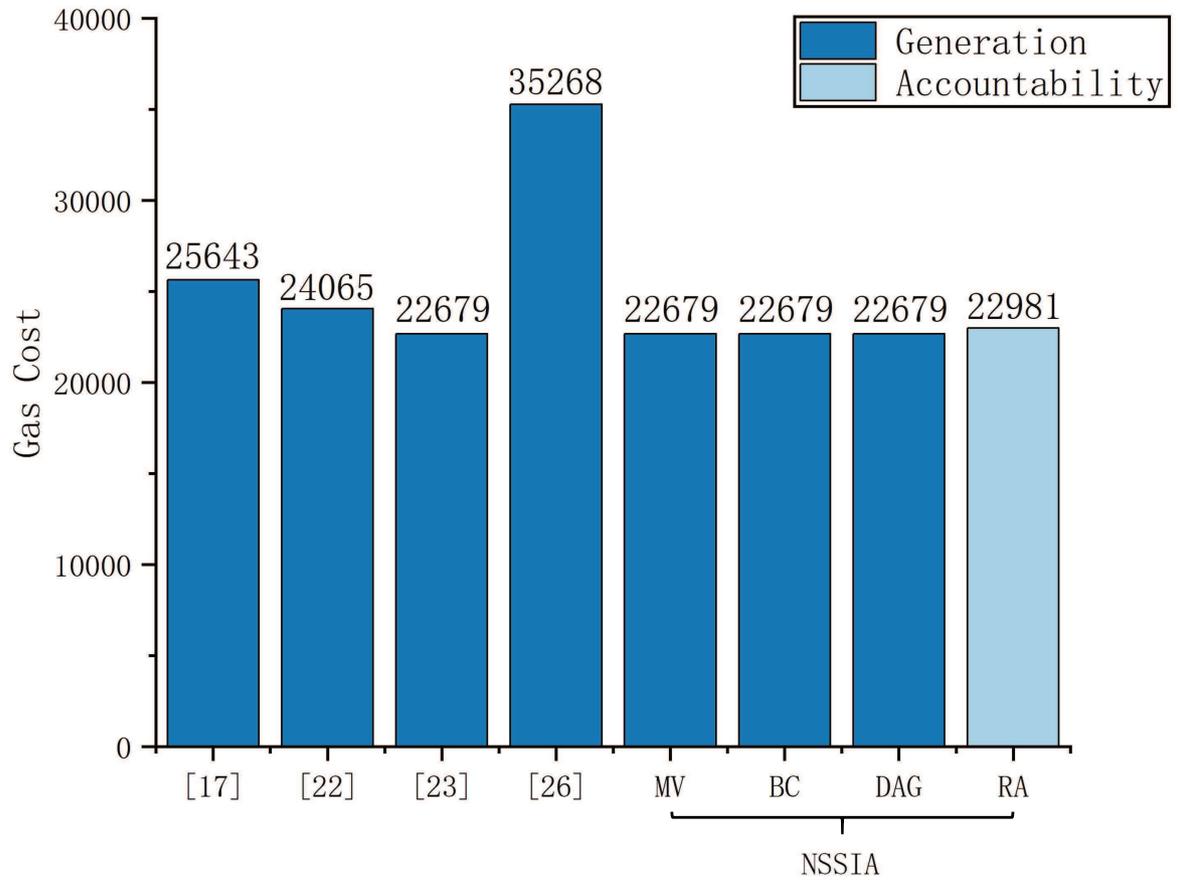}
    \end{center}
  \caption{Comparison of Gas cost when writing data\label{fig7}}
\end{figure}

Writing and reading are the main interactions between entities and the consortium blockchain. 
Therefore, we record the proof data by deploying smart contract on Rinkeby and count the Gas spent on contract deployment and invocation. 
And, since Maram et al. \cite{maram2021candid} has no blockchain deployed, we compare our scheme with the above SSI schemes \cite{takemiya2018sora,lee2019sims,zheng2019blockchain,bandara2021blockchain} based on the data in Section~\ref{subsec6.3}, as shown in Fig~\ref{fig7}. 

As we can see from Figure~\ref{fig7}, during the generation phase, there are more than $200$ bytes of data recorded on the blockchain in \cite{takemiya2018sora}, which costs approximately $25643$ Gas. 
In \cite{lee2019sims}, a total of $102$ bytes of data are written to the blockchain and the cost is $24065$ Gas. 
Zheng et al. \cite{zheng2019blockchain} records $20$ bytes of certificates on the blockchain, costing $22679$ Gas. 
The cost of recalling the contract to write $800$ bytes of identity proof in \cite{bandara2021blockchain} is $35268$ Gas. 
In the phase~\ref{subsec4.1}, the contract in our NSSIA is deployed and the cost is $176335$ Gas. 
Unlike the above schemes where only one entity interacts with the blockchain, in our scheme, different entities are responsible for interacting with the blockchain at different stages of identity generation descripted in Section~\ref{sec4}. 
Concretely, in the digitization stage, metadata verifier (MV) and biometric collector (BC) respectively record 20-byte metadata and biometric proofs into the blockchain, with an overhead of $22679$ Gas. 
And, in the generation stage, the proof of the generated digital avatar-i (DA) is written into the blockchain by digital avatar-i generator (DAG), which also costs $22679$ Gas. 
Although the total cost is $22679*3 = 68037$ Gas which is greater than other schemes, the Gas cost of our scheme is actually the lowest due to the spread over different entities. 
In addition, we greatly reduce the risk of centralization of power compared to other schemes. 

In terms of accountability, the overhead of the above schemes is $0$ due to the lack of accountability mechanism. 
In our scheme, there are $34$ bytes of log information recorded by RA on the blockchain, and the cost is $22981$ Gas, which is almost the lowest compared with the overhead of the generation phase. 
All in all, regardless of the generation stage or the accountability stage, the Gas cost of our scheme is not prohibitive for practice use. 
Furthermore, the decentralization of regulatory authorities' power guarantees fair audit and protects users privacy. 
\section{Conclusion and future work}\label{sec7}
A new self-sovereign identity scheme with accountability is proposed in this paper, where the executable code is introduced to allow each user to independently control their own identity, referred as the digital avatar-i (DA), and malicious users can be fairly regulated without violating the privacy of legitimate ones. 
For concreteness, one and only individual-specific executable code is generated for each user to interact with others in metaverse without a third-party program, in which biometrics are integrated into the code to enhance uniqueness and user control. 
The hash of the individual-specific executable code is used as an identifier and each user can store, read and prove identities with service providers through his/her own local executable code. 
Furthermore, a joint accountability mechanism is introduced to balance the privacy and accountability, where shamir(t, n) threshold algorithm is used to decentralize the power of each regulatory authority and hide users' information in reality, and the impartial audit is further guaranteed by a consortium blockchain. 
The security analysis illustrates that our NSSIA can resist multiple security threats such as sybil attacks, impersonation attacks and so on. 
And the analysis result on SSI properties shows that we have satisfied all the six SSI properties in identity generation phase. 
Compared with the state-of-the-art schemes, the extensive experiment results in performance indicates that the overhead of our NSSIA is not unreasonable for practical use. 

For future work, we will pay attention to the difficulties existing in the use of the DA, such as unlinkability, right to be forgotten, etc., and the full design of Section~\ref{subsec4.4} will be presented. 
Meanwhile, striking a balance between privacy and accountability when using the DA to interact with others in cyberspace is also the focus of our research. 

\bibliographystyle{unsrt}
\bibliography{NSSIA}

\end{document}